\documentclass[12pt]{article}
\usepackage[english]{babel}
\usepackage[T1]{fontenc}
\usepackage{amsmath}
\usepackage{amsfonts}

\newlength{\dinwidth}
\newlength{\dinmargin}
\newcommand{\czn}{C_o^{\infty}(\mathbb{R}^3)}
\newcommand{\cznn}{C_o^{\infty}(\mathbb{R}^4)}
\newcommand{\cz}{C_o^{\infty}}
\newcommand{\mco}{\mathcal{O}}
\newcommand{\mfa}{\mathfrak{A}}
\newcommand{\spt}{\mathbb{R}^4}
\newcommand{\hil}{\mathcal{H}}
\newcommand{\cpo}{P_{+}^{\uparrow}}
\newcommand{\vac}{\Omega}
\newcommand{\ft}{\tilde{f}}
\newcommand{\fr}[2]{\frac{#1}{#2}}
\newcommand{\Ga}{\Gamma}
\newcommand{\hti}{\tilde{h}}
\newcommand{\vx}{\vec{x}}
\newcommand{\vep}{\vec{p}}

\newcommand{\og}{\omega(\vep)}

\newcommand{\eps}{\epsilon}

\newcommand{\A}[1]{A_{#1}(f_{#1T})}
\newcommand{\As}[1]{A_{#1}(f_{#1T}^s)}
\newcommand{\Ah}[1]{\widehat{A}_{#1}(\widehat{f}_{{#1}T}^{\hat{s}})}
\newcommand{\Ahh}[1]{\widehat{A}_{#1}(\widehat{f}_{{#1}T})}
\newcommand{\An}[1]{A_{#1}(f_{#1})}
\newcommand{\Ahhn}[1]{\widehat{A}_{#1}(\widehat{f}_{{#1}})}

\newcommand{\At}[1]{A_{#1}(t_{#1},\vx_{#1})}
\newcommand{\vu}{\vec{u}}
\newcommand{\si}{\sigma}
\newcommand{\hf}{\hat{f}}
\newcommand{\cf}{\check{f}}

\def\proof{\noindent{\bf Proof. }}
\def\qed{$\Box$\medskip}

\newtheorem{theoreme}{Theorem } [section]
\newtheorem{proposition}[theoreme]{Proposition}
\newtheorem{lemma}[theoreme]{Lemma}
\newtheorem{definition}[theoreme]{Definition}
\newtheorem{corollary}[theoreme]{Corrolary}
\newtheorem{remark}[theoreme]{Remark}
\newtheorem{example}[theoreme]{Example}
\newcommand{\beq}{\begin{equation}}
\newcommand{\eeq}{\end{equation}}
\newcommand{\beqa}{\begin{eqnarray}}
\newcommand{\eeqa}{\end{eqnarray}}
\newcommand{\ben}{\begin{arabicenumerate}}
\newcommand{\een}{\end{arabicenumerate}}
\newcommand{\bex}{\begin{example}}
\newcommand{\eex}{\end{example}}
\newcommand{\ber}{\begin{remark}}
\newcommand{\eer}{\end{remark}}
\newcommand{\bec}{\begin{corollary}}
\newcommand{\eec}{\end{corollary}}
\newcommand{\bep}{\begin{proposition}}
\newcommand{\eep}{\end{proposition}}
\def\bel{\begin{lemma}}
\def\eel{\end{lemma}}
\def\bet{\begin{theoreme}}
\def\eet{\end{theoreme}}
\def\bed{\begin{definition}}
\def\eed{\end{definition}}

\begin{document}
\title{Haag-Ruelle scattering theory in presence of massless particles}
\author{{Wojciech Dybalski\footnote{e-mail: dybalski@theorie.physik.uni-goettingen.de}}\\ [5mm] 
  Institut f\"ur Theoretische Physik, Universit\"at G\"ottingen, \\[2mm]
Friedrich-Hund-Platz 1, D-37077 G\"ottingen - Germany }
\date{}
\maketitle
\begin{abstract}
Within the framework of local quantum physics we construct a scattering theory of
stable, massive particles without assuming mass gaps. This extension of the Haag-Ruelle
theory is based on advances in the harmonic analysis of local operators.
Our construction is restricted to theories complying with a regularity property
introduced by Herbst. The paper concludes with a brief discussion of the status of this
assumption.   
\end{abstract}

%
\section{Introduction}
The physical interpretation of relativistic quantum field theories is primarily based
on collision theory which has been a fundamental issue for more than five decades.
Whereas collision theory in the purely massive case is well understood by the work of
H. Lehmann, K. Symanzik and W. Zimmerman on one hand \cite{LSZ} and by R. Haag
and D. Ruelle on the other \cite{Ha,Ru}, the situation is less clear in theories with long range
forces. There collision theory is under complete control only for massless particles 
\cite{Buch2,Buch4}; yet a general method for the treatment of collisions of stable massive particles
is missing to date in these cases, not to speak about the so-called infraparticle problem 
in the presence of Gauss' law \cite{Sch,Buch5}. Those difficulties manifest themselves
already at the level of the classical Maxwell-Dirac system in the definition of wave
operators \cite{FST}.

It is the aim of the present article to prove that the Haag-Ruelle collision theory can be 
extended to stable massive particles obeying a sharp dispersion law in the presence of massless excitations.
Thus we do not touch upon the infraparticle problem, but our arguments are applicable, for
example, to electrically neutral, stable particles such as atoms in quantum electrodynamics.
Before we enter into this discussion we briefly outline our notation, state our assumptions and comment on previous approaches to this problem.       

To keep the notation simple, we will consider the scattering of a single 
type of massive Bosons in the presence of massless Bose particles. 
We base our theory on a net $\mco\to\mfa(\mco)$ of local C*-algebras
attached to open, bounded regions $\mco\subset\spt$.  The
global algebra $\mfa$ of the net is assumed to act irreducibly
on the Hilbert space $\hil$. We further suppose that 
$\mfa(\mco_1)\subset\mfa(\mco_2)^{'}$ if $\mco_1\subset \mco_2^{'}$,
where $\mco_2^{'}$ is the spatial complement of $\mco_2$ and a prime
over an algebra denotes its commutant.
Moreover, $\hil$ carries a continuous unitary representation
$L\to U(L)$ of the covering group of the Poincar\'e group $\cpo$ 
such that:
\beq 
U(L) \mfa(\mco) U(L)^{-1}=\mfa(L\mco).
\eeq
The joint spectrum of the generators of translations $P^{\mu}$ is 
contained in the forward light cone. The vacuum is modelled by a unique 
(up to a phase) unit vector $\vac$ in $\hil$, which is invariant under 
all $U(L)$,  $L\in\cpo$. A single massive particle is described by a state in 
a subspace $\hil_1\subset\hil$ on which the $U(L)$ act like an irreducible 
representation of $\cpo$ with mass $m>0$. We denote the spectral measure of
the energy-momentum operators by $E$ and the projection on $\hil_1$ by $E_m$. 
In the pioneering work of Haag \cite{Ha} and Ruelle \cite{Ru} these 
general postulates were amended by two additional requirements:
\begin{enumerate}
\item [A.] The time-dependent operators $A(f_T)=\int A(x)f_T(x) d^4x$, 
constructed from $A(x)=U(x)AU(x)^{-1}$, $A\in \mfa(\mco)$ and suitably chosen sequences 
of functions $f_T\in S(\spt)$, satisfy $A(f_T)\vac\neq 0$, $A(f_T)\vac\in\hil_1$ and $\fr{d}{dT}A(f_T)\vac=0$. 
\item [M.] The vacuum is isolated from the rest of the energy-momentum spectrum.
 \end{enumerate}
Both of these conditions are ensured if the mass $m$ is an isolated eigenvalue of the mass operator $\sqrt{P^{\mu}P_{\mu}}$. On the other hand, if the mass
of the particle in question is an embedded eigenvalue then it seems difficult to
meet the requirement A. It was, however, noticed  
by I. Herbst \cite{He} that, in fact, it is only needed in the proof that 
$s-\lim_{T\to\infty}A(f_T)\vac$ is a non-zero vector in $\hil_1$ and 
$\|\fr{d}{dT}A(f_T)\vac\|$ is an integrable function of $T$. 
We summarize here Herbst's analysis since it will be the starting point of our 
considerations: The operators $A(f_T)$ are constructed in a slightly different manner 
than in the work of Haag and Ruelle: 
First, a local operator $A$ is smeared in space with a regular solution of the Klein-Gordon equation 
$f(t,\vx)=\fr{1}{(2\pi)^3} \int e^{-i\og t+i\vep\vx}\ft(\vep)d^3p $, (where $\ft(\vep)\in\czn$, 
$\og=\sqrt{\vep^2+m^2}$) :
\beq
A_t(f)=\int A(t,\vx)f(t,\vx)d^3x. \label{fTdef1}
\eeq
Next, to construct the time averaging function, we choose $s(T)=T^{\nu}$, 
$0<\nu<1$ and a positive function $h\in S(\mathbb{R})$ such that its Fourier transform satisfies $\hti\in\cz (\mathbb{R})$, $\hti(0)=1$. Then we set $h_T(t)=\fr{1}{s(T)}h(\fr{t-T}{s(T)})$ and define \cite{He,Buch2}:
\beq
A(f_T)=\int h_T(t) A_t(f) dt. \label{fTdef2}
\eeq 
It is clear from the formulas (\ref{fTdef1}) and (\ref{fTdef2}) that $f_T(x)=h_T(x^0)f(x^0,\vec{x})$.  
Its Fourier transform $\ft_T$ has a compact support which approaches a compact subset of the
mass hyperboloid as $T\to\infty$. In view of this fact we will refer to $A(f_T)$ as creation operators
and to $A(f_T)^*$ as annihilation operators. This terminology is also supported by the following
simple calculation:
\beq
s-\lim_{T\to\infty}A(f_T)\vac=E_mA(f)\vac,
\eeq
where $A(f)=A_{t=0}(f)$. The integrability condition requires the following assumption:
\begin{enumerate}
\item[A'.] There exist operators $A\in\mfa(\mco)$ such that $E_mA\vac\neq 0$ and, for every $\delta\geq 0$,
\beq
\|E(m^2-\delta\leq P^2\leq m^2+\delta)(1-E_m)A\vac\|\leq c\,\delta^\epsilon
\label{Regularity1}
\eeq
for some $c,\epsilon>0$. We refer to such operators as 'regular'.
\end{enumerate}
For regular operators there follows a bound:
\beq
\|\fr{d}{dT}A(f_T)\vac\|\leq\fr{c}{s(T)^{1+\eps}} \label{derivative}
\eeq
which implies integrability if $\nu>\fr{1}{1+\eps}$. Now we are ready to state the main 
result of Herbst; we restrict attention to the outgoing asymptotic states $\Psi^+$, since 
the case of incoming states is completely analogous.
\bet\cite{He}\label{Herbsttheorem}  Suppose that the theory respects the conditions M and A'.
Then, for regular operators $A_i$, $i=1\ldots n$, there exists the limit:
\beq
 \Psi^{+}=s-\lim_{T\to\infty} A_1(f_{1T})\ldots A_n(f_{nT})\vac
\eeq
and it depends only on the single-particle states $E_mA_i(f_i)\vac$. 
Moreover, given two states $\Psi^{+}$ and $\hat\Psi^{+}$ constructed as above using creation
operators $\A{i}$ and $\Ahh{i}$, their scalar product can be calculated as follows: 
 \beq 
(\Psi^{+}|\widehat\Psi^{+})=
\sum_{\si\in S_n}(\vac|\An{1}^*E_m\Ahhn{\si_{1}}\vac)\ldots(\vac|\An{n}^*E_m\Ahhn{\si_{n}}\vac).
\eeq
Here the sum is over all permutations of an n-element set.
\eet
It was, however, anticipated already by Ruelle \cite{Ru} that in a purely massive theory the condition A can be replaced by the following, physically meaningful, stability requirement:
 \begin{enumerate}
 \item[S.] In a theory satisfying M a particle can only be stable if, in its 
 superselection sector, its mass is separated from the rest of the spectrum by a
 lower and upper mass gap.   
 \end{enumerate}
This condition is also stated in Herbst's work \cite{He}, but he expects that a scattering
theory can be a necessary tool to study the superselection structure.      
Subsequent analysis by D. Buchholz and K. Fredenhagen \cite{Buch1} clarified this issue:
There exist interpolating fields which connect the vacuum with the sector of the
given particle. Although they are, in general, localized in spacelike cones, they
can be used to construct a collision theory. Thereby there exists a prominent alternative to
the approach of Herbst in the realm of massive theories.  

It is the purpose of our investigations to extend Herbst's result to the situation when 
massless particles are present, that is the conditions M and S do not hold. 
A model physical example of a system with a sharp mass immersed in a spectrum
of massless particles is the hydrogen atom in its ground state from the point of view 
of quantum electrodynamics.  
Although the approach of Herbst seems perfectly adequate to study such situations, the 
original proof of Theorem \ref{Herbsttheorem} does not work because of the slow, 
quadratic decay of the correlation functions. In order to overcome this difficulty 
we apply the novel bounds on creation operators obtained by D. Buchholz \cite{Buch3}.
Namely, if $\Delta$ is any compact subset of the energy-momentum spectrum and 
$\ft$ vanishes sufficiently fast at zero then: 
\beqa
\|A(f_T)E(\Delta)\|\leq c  \label{bound1}\\
\|A(f_T)^*E(\Delta)\|\leq c \label{bound2}
\eeqa
where the constant $c$ does not depend on time. 

Our paper is organized as follows: In Section 2 we prove the existence of asymptotic states 
and verify that the limits are independent of the
actual value of the parameter $0<\nu<1$ chosen in the time averages of the operators $A(f_T)$. 
This property allows us to apply in Section 3 the methods from
the collision theory of massless Bosons \cite{Buch4} in order to calculate the scalar  
product of asymptotic states. In the Conclusion we summarize our results and discuss 
the status of the condition A'.

\section{Existence of Asymptotic States}
In order to prove the existence of asymptotic states we need information about 
the time evolution of operators $A(f_T)$ and their commutators. It is the purpose
of the two subsequent lemmas to summarize the necessary properties:  

\bel \label{norm} Let $A(f_T)^{\#}$ denote $A(f_T)$ or $A(f_T)^*$. Then:
\begin{enumerate}
\item[a)] $\|A(f_T)^{\#}\|\leq cT^{3/2}.$ 
\item[b)] $E(\Delta_1)A(f_T)^{\#}E(\Delta_2)=0$ if 
$\Delta_1\cap(\Delta_2\pm\textrm{supp}\ft_T)=\emptyset$. The $(+)$ sign holds for
$A(f_T)$, $(-)$ for $A(f_T)^*$.
\item[c)] Suppose that the functions $\ft_i$, $i=1\ldots n$, vanish sufficiently fast at zero. 
Then, for any compact subset $\Delta$ of the energy-momentum spectrum:
 \beq
 \|\A{1}^{\#}\ldots \A{n}^{\#}E(\Delta)\|\leq c_1. \label{multiplebound}
 \eeq
\end {enumerate}
The constants $c$, $c_1$ do not depend on $T$.
\eel  
\proof
\begin{enumerate}
\item[a)] The statement follows from the estimate:
\begin{eqnarray}
\|A(f_T)^{\#}\|&\leq&\|A^{\#}\|\int dt h_T(t)\int d^3x|f(t,\vx)|
\leq c_0\int dt h_T(t)(1+|t|)^{3/2}\nonumber\\
&=& c_0\int dt h(t)(1+|s(T)t+T|)^{3/2}
\leq c T^{3/2},
\end{eqnarray}
where in the second step we used the properties of regular solutions 
of the Klein-Gordon equation \cite{Ru}.
\item[b)] See, for example, \cite{Ha1}. 
\item[c)] For $n=1$ the assertion follows from $(\ref{bound1})$ and $(\ref{bound2})$.
Assuming that (\ref{multiplebound}) is valid for $n-1$ and making use of
part b of this lemma we estimate:
\beqa
\|\A{1}^{\#}\ldots \A{n}^{\#}E(\Delta)\|\nonumber\\
=\|\A{1}^{\#}\ldots\A{n-1}^{\#}E(\Delta\pm\textrm{supp}\ft_{nT}) \A{n}^{\#}E(\Delta)\|\nonumber\\
\leq\|\A{1}^{\#}\ldots\A{n-1}^{\#}E(\Delta\pm\textrm{supp}\ft_{nT})\|\|\A{n}^{\#}E(\Delta)\|.
\eeqa 
The last expression is bounded by the inductive assumption and the support properties
of functions $\ft_T$.   
\end {enumerate} \qed\\
Now we turn our attention to the commutators of the operators $A(f_T)$.
It will simplify this discussion to decompose the function $f_T(x)$ into
its compactly supported dominant contribution and a spatially extended,
but rapidly decreasing remainder \cite{PFG}. To this end, let us define the velocity
support of the function $\ft$: 
\beq
\Ga(\ft) =\{(1,\fr{\vep}{\og}): \vep\in \mathrm{supp}\ft \}.
\eeq
We introduce a function $\chi_{\delta}\in\cznn$ such that $\chi_{\delta}=1$ on $\Ga(\ft)$
and $\chi_{\delta}=0$ in the complement of a slightly larger set $\Ga(\ft)_{\delta}$. 
$\hat{f}_T(x):=f_T(x)\chi_{\delta}(x/T)$ is the asymptotically dominant part of
$f_T(x)$, whereas $\check{f}_T(x):=f_T(x)(1-\chi_{\delta}(x/T))$ tends rapidly to 
zero with $T\to\infty$ \cite{He,Hepp2}. In particular, for
each natural $N$ and some fixed $N_0>4$ there exists a constant $c_N$ such that:
\beq
\int |\check{f}_T(x)|d^4x\leq c_N\fr{s(T)^{N+N_0}}{T^N}. \label{integral}
\eeq    
We remark that this bound relies on the slow increase of the function $s(T)$, 
so the condition A' cannot be eliminated simply by modifying this function. 

As was observed first by Hepp \cite{Hepp1}, 
particularly strong estimates on commutators can be obtained in the case of particles moving with different velocities:       
\bel \label{com} Let $A_1(f_{1T})$, $A_2(f_{2T})$, $A_3(f_{3T})$ be defined as above. 
Moreover, let $\ft_1$, $\ft_2$ have disjoint velocity supports. Then, for each natural $N$, there
exists a constant $c_N$ such that:
 \begin {enumerate}
 \item[a)] $\|[A_1(f_{1T}),A_2(f_{2T})]\|\leq \fr{c_N}{T^N}.$
 \item[b)] $\|[A_1(f_{1T}),[A_2(f_{2T}),A_3(f_{3T})]]\|\leq\fr{c_N}{T^N}.$
\end{enumerate}
 The same estimates are valid if some of the operators $A(f_T)$ are replaced
 by their adjoints or time derivatives.
  \eel
\proof
\begin {enumerate}
\item[a)]  We make use of the decomposition: $f_T=\hf_T+\cf_T$:
 \beqa
 [A_1(f_{1T}),A_2(f_{2T})]
 =[A_1(\hf_{1T}),A_2(\hf_{2T})]+[A_1(\hf_{1T}),A_2(\cf_{2T})] \nonumber\\
 +[A_1(\cf_{1T}),A_2(\hf_{2T})]+[A_1(\cf_{1T}),A_2(\cf_{2T})].
 \eeqa
The first term on the r.h.s. is a commutator of two local operators. For
sufficiently large $T$ their localization regions become spatially separated
because of disjointness of the velocity supports of $\ft_1$ and $\ft_2$. Then the
commutator vanishes by virtue of locality. Each of the remaining terms contains
a factor $A(\cf_T)$ which decreases in norm faster than any power of $T^{-1}$
by the estimate (\ref{integral}). It is multiplied by $A(\hf_T)$ which increases
in norm only as $T^{3/2}$ by Lemma \ref{norm} a.

\item[b)] First, let us suppose that $\ft_3$ and $\ft_2$ have disjoint velocity supports. Then $[A_2(f_{2T}),A_3(f_{3T})]$ decreases fast in norm as a consequence 
of part a of this Lemma. Recalling that the norm of $A_1(f_{1T})$ increases only as 
$T^{3/2}$ the assertion follows. Now suppose that $\ft_3$ and $\ft_1$ have disjoint velocity supports. 
Then, by application of the Jacobi identity, we arrive at the previous situation.   
In the general case we use a smooth partition of unity to decompose $\ft_3$ into a
sum of two functions, each belonging to one of the two special classes studied
above. 
\end {enumerate}
The statement about adjoints is obvious. To justify the claim concerning  
derivatives we note that:
\beq
\fr{d}{dT}A(f_T)=-\fr{1}{s(T)}\bigg\{\fr{ds(T)}{dT}\big(A(f_T)+A(f_{aT})\big)+A(f_{bT})\bigg\}, 
\label{derivative1}
\eeq
where $f_{aT}$ is constructed using $h_a(t)= t\fr{dh(t)}{dt}$ and $f_{bT}$ 
contains $h_b(t)=\fr{dh(t)}{dt}$. Although $h_a(t)$ and $h_b(t)$ do not satisfy 
all the conditions imposed previously on functions $h(t)$, they are elements of 
$S(\mathbb{R})$. This property suffices to prove the decomposition 
$f_T=\hat{f}_T+\check{f}_T$\qed\\
After having constructed creation operators and studied their properties, it
will be fairly simple to demonstrate the existence of asymptotic
states. The following theorem uses the original method of Haag \cite{Ha} 
modified by Araki \cite{Ara}:
\bet\label{Haag}
Suppose that local operators $A_1,\ldots A_n$ are regular, $\ft_1,\ldots \ft_n$ have
disjoint velocity supports and vanish sufficiently fast at zero. Moreover, $s(T)=T^{\nu}$,
$\fr{1}{1+\eps}<\nu<1$, where $\eps$ is the exponent appearing in the regularity condition A'. 
Let us denote $\Psi(T)=A_1(f_{1T})\ldots A_n(f_{nT})\vac$. Then there exists the limit 
$\Psi^+=s-\lim_{T\to\infty}\Psi(T)$ and it is called an asymptotic state. 
\eet
\proof
We verify the Cauchy condition using Cook's method:
\beq
\|\Psi(T_2)-\Psi(T_1)\|\leq \int_{T_1}^{T_2}\|\fr{d\Psi(T)}{dT}\|dT.\label{Cauchy}
\eeq
It now has to be checked whether the integrand decays sufficiently
fast when $T\to\infty$. By using the Leibniz rule, and then commuting the derivatives of creation operators with the other operators until they act on the vacuum, we arrive at the 
following expression:
\beqa
\fr{d\Psi}{dT}&=&\sum_{k=1}^{n} \A{1}\ldots \fr{d}{dT}\A{k}\ldots \A{n}\vac\nonumber\\
&=&\sum_{k=1}^{n}\big\{\sum_{l=k+1}^{n}\A{1}\ldots [\fr{d}{dT}\A{k},\A{l}]\ldots\A{n}\vac\nonumber\\
	&+&\A{1}\ldots\check{k}\ldots\A{n}\fr{d}{dT}\A{k}\vac\big\},
\eeqa
where $\check{k}$ denotes omission of $A_k(f_{kT})$.
Each term containing commutators vanishes in norm faster than any power of 
$T^{-1}$. It follows from the fact that the rapid decay of commutators, proved
in Lemma \ref{com} a, suppresses the polynomial increase of $\|A(f_T)\|$ shown in
Lemma \ref{norm} a. To estimate the 
remaining terms we first note that, by virtue of the formula (\ref{derivative1}) 
and Lemma \ref{norm} b, the vector $\fr{d}{dT}\A{k}\vac$ has a compact spectral support
$\Delta$. Consequently:
\beqa
 &  & \|\A{1}\ldots\check{k}\ldots\A{n}\fr{d}{dT}\A{k}\vac\|  \nonumber\\
 &= & \|\A{1}\ldots\check{k}\ldots\A{n}E(\Delta)\fr{d}{dT}\A{k}\vac\| \nonumber\\
 &\leq& \|\A{1}\ldots\check{k}\ldots\A{n}E(\Delta)\|\|\fr{d}{dT}\A{k}\vac\|
 \leq \fr{c}{s(T)^{1+\eps}},
\eeqa
where in the last step we made use of Lemma \ref{norm} c and the estimate 
(\ref{derivative}). As $\nu(1+\eps)>1$, the integral (\ref{Cauchy}) 
tends to zero when $T_1,T_2\to\infty$ and the Cauchy condition is satisfied.
\qed\\   
It is a remarkable feature of asymptotic states with disjoint velocity supports
that already at this stage it is possible to prove that they depend only
on the single-particle states $E_mA(f)\vac$ rather than on the
specific $A$, $\ft$, $h$, and $s$ that were used to construct them. As the possibility
to relax the increase of functions $s(T)$ is particularly important for us, we 
temporarily introduce the notation $A(f_T^s)$ to distinguish between operators
containing different functions $s(T)$.  
\bel\cite{Goettingen}\label{sT}
Suppose that the families of operators $\As{1}\ldots \As{n}$, resp. 
$\Ah{1}\ldots \Ah{n}$,  satisfy the following conditions:
\begin{enumerate} 
\item[a)] The functions $\ft_1\ldots\ft_n$, resp. $\widehat\ft_1\ldots\widehat\ft_n$, 
have, within each family, disjoint velocity supports and vanish sufficiently fast at zero.  
\item[b)] $E_m A_i(f_i)\vac=E_m\widehat{A}_i(\widehat{f_i})\vac$, $i=1\ldots n$, i.e. the single-particle 
states corresponding to the two families of operators coincide.
\item[c)] $\Psi^{+}=s-\lim_{T\to\infty}A_{1}(f_{1T}^s)\ldots A_{n}(f_{nT}^s)\vac$
exists.  
\end{enumerate}
Then the limit $\widehat\Psi^{+}=s-\lim_{T\to\infty}\Ah{1}\ldots \Ah{n}\vac$ exists and
coincides with $\Psi^{+}$.
\eel
\proof
We proceed by induction. For $n=1$ the assertion is satisfied by assumption.
Let us assume that it is satisfied for states involving $n-1$ creation operators.
Then the following inequality establishes the strong convergence of the net $\As{1}\Ah{2}\ldots \Ah{n}\vac$: 
\beqa
\|\As{1}\As{2}\ldots \As{n}\vac-
\As{1}\Ah{2}\ldots \Ah{n}\vac\|\nonumber\\ 
\leq \|\As{1}E(\Delta)\| \|\As{2}\ldots \As{n}\vac-\Ah{2}\ldots \Ah{n}\vac\|,
\eeqa
where $\Delta$ is the spectral support of the product of creation operators acting on the vacuum which  
is compact by Lemma \ref{norm} b. The r.h.s. of this expression vanishes in the 
limit of large $T$ as a consequence of the estimate (\ref{bound1}) and the induction hypothesis. 
By applying the bound on commutators proved in Lemma \ref{com} a and the estimate from 
Lemma \ref{norm} a we verify that also $\Ah{2}\ldots \Ah{n}\As{1}\vac$ converges strongly and has
the same limit. Finally, our claim follows from the estimate:
\beqa
\|\Ah{2}\ldots \Ah{n}(\Ah{1}-\As{1})\vac\|\nonumber\\
\leq \|\Ah{2}\ldots \Ah{n}E(\Delta_1)\|\|(\Ah{1}-\As{1})\vac\|,
\eeqa
where $\Delta_1$ is again a compact spectral support.
The r.h.s. of this inequality tends to zero with $T\to\infty$ by our assumption
concerning single-particle states and the bound in Lemma \ref{norm} c.\qed\\
\section{Fock Structure of Asymptotic States}
It was instrumental in the original proof of the existence of asymptotic states that $s(T)=T^\nu$, where
$\nu$ was sufficiently close to one. Lemma \ref{sT} allows us to relax this condition and choose any
$0<\nu<1$. Using this piece of information we will verify the Fock structure of the
scattering states by the following strategy: First, we establish a counterpart of the relation
$a a^*\vac=(\vac|a a^*|\vac)\vac$ satisfied by ordinary creation and annihilation operators.
Once this equality is proven in the sense of strong limits, we combine it with
the double commutator bound from Lemma \ref{com} b to obtain the factorization of 
the scalar product of scattering states. 

 We start from two definitions: $\mathcal{C}_R$ is the double cone given by the intersection of
the forward cone with tip in $(-R,0)$ and the backward cone with tip in $(R,0)$.  
By $\mfa_0$ we denote the weakly dense subalgebra of $\mfa$ consisting of operators 
for which the operator valued functions $ x\to A(x)$ are infinitely often differentiable 
in the norm topology. (We remark that, given any regular operator, we can construct
a regular operator in $\mfa_0$ by smearing it with a smooth function.)  

We will benefit from an estimate obtained in \cite{Buch4} to study the scattering theory
of massless Bosons. It was derived combining geometrical considerations with the 
result due to Araki, Hepp and Ruelle on the quadratic decay of the two-point function of 
suitable local operators \cite{AHR}. We state it in a form adapted to our situation:

\bel
Let $A_1\ldots A_4\in\mfa_0$ be localized in double cones $\mathcal{C}_{R_1}\ldots\mathcal{C}_{R_4}$. We define:
\beq
K=(\vac|[A_1(t_1,\vx_1),A_2(t_2,\vx_2)](1-E_0)[\At{3},\At{4}]\vac).
\eeq
Then the following estimate holds:
\beq
|K|\leq c\chi(|\vx_1-\vx_2|\leq R) \chi(|\vx_3-\vx_4|\leq R)\cdot\left\{\begin{array}{cc}
 1 & \textrm{if $|\vx_2-\vx_3|\leq 4R$}\\
\fr{R^3}{|\vx_2-\vx_3|^2+R^2} & \textrm{if $|\vx_2-\vx_3|>4R$}
\end{array}\right. 
\eeq
where $R=\sum_{i=1}^4(R_i+|t-t_i|)$, $t=\fr{1}{4}(t_1+t_2+t_3+t_4)$, 
$\chi$ are the characteristic functions of the respective sets. The constant $c$ depends neither on 
$t_1\ldots t_4$ nor on $R_1\ldots R_4$. 
 \eel
Now we are ready to prove that a product of a creation and annihilation operator acting on the vacuum reproduces it:
\bep
Suppose that $A_1$,$A_2$ are local operators on $\mfa_0$, $s(T)=T^{\nu}$, $\nu<1/8$. Then
\beq
s-\lim_{T\to\infty}\A{1}^*\A{2}\vac=(\vac|\An{1}^*E_m\An{2}\vac)\vac. 
\eeq
\eep
\proof
We start by performing the integration of the function $K$ from the preceding Lemma 
with the regular Klein-Gordon wave-packets 
and estimating the behaviour of the resulting function of $t_1\ldots t_4$. We will change the variables
of integration to $\vu_1=\vx_2-\vx_1$, $\vu_2=\vx_2$, $\vu_3=\vx_3-\vx_2$, $\vu_4=\vx_4-\vx_3$.
In the region $|\vx_2-\vx_3|> 4R$ we obtain:
\beqa
&  &\int d^3x_1|f_1(t_1,\vx_1)|\ldots \int d^3x_4|f_4(t_4,\vx_4)| |K|\nonumber\\
&\leq& cR^9(1+|t_1|)^{-3/2}(1+|t_4|)^{-3/2}\int d^3u_2\int d^3 u_3
\fr{|f_2(t_2,\vu_2)||f_3(t_3,\vu_3+\vu_2)|}{|\vu_3|^2+R^2}\nonumber\\
&\leq& cR^9(1+|t_1|)^{-3/2}(1+|t_4|)^{-3/2}(1+|t_2|)^{3/2}.\quad\label{integrals}
\eeqa
In the last step we used the inequality  $2\fr{|f_3(\cdot,\cdot)|}{|\vu_3|^2+R^2}\leq 
(\fr{1}{|\vu_3|^2+R^2})^2+|f_3(\cdot,\cdot)|^2$ in order to verify that the integral over $\vu_3$
is bounded in $t_3$. In the region  $|\vx_2-\vx_3|\leq 4R$ our estimate becomes improved by the
factor $(1+|t_3|)^{-3/2}$, so (\ref{integrals}) holds without restrictions.

Since $\A{}^*\vac=0$ for sufficiently large $T$, we can estimate:
\beqa
& &\|\A{1}^*\A{2}\vac-(\vac|\A{1}^*\A{2}\vac)\vac\|^2\nonumber\qquad\\
&=&(\vac|[\A{2}^*,\A{1}](1-E_0)[\A{1}^*,\A{2}]\vac)\nonumber\\
&\leq&\int dt_1 h_T(t_1)\ldots\int dt_4 h_T(t_4)
cR^9(1+|t_1|)^{-3/2}(1+|t_4|)^{-3/2}(1+|t_2|)^{3/2}\nonumber\\
&\leq& C\fr{s(T)^{12}}{T^{3/2}}.
\eeqa
Now the assertion follows from the slow increase of the function $s(T)$.
\qed\\
After this preparation it is straightforward to calculate the scalar product of two asymptotic
states. 
\bet Suppose that  $\A1\ldots\A{n}$, resp. $\Ahh1\ldots\Ahh{n}$, are two families 
of creation operators constructed from local operators on $\mfa_0$, functions
 $\ft_i$, resp. $\widehat\ft_i$, $i=1\ldots n$, vanishing sufficiently fast at zero 
and having, within each family, disjoint velocity supports. Moreover, $s(T)=T^{\nu}$, where  
$0<\nu<1$. Then: 
\beqa
\lim_{T\to\infty}
(\vac|\A{n}^*\ldots\A1^*\Ahh1\ldots\Ahh{n}\vac)\nonumber\\
=\sum_{\si\in S_n}(\vac|\An{1}^*E_m\Ahhn{\si_{1}}\vac)\ldots(\vac|\An{n}^*E_m\Ahhn{\si_{n}}\vac).
\eeqa
Here the sum is over all permutations of an n-element set.
\eet
\proof
First, we make use of Lemma \ref{sT} to ensure a sufficiently slow
increase of the function $s(T)$. Next,  
we proceed by induction. For $n=1$ the theorem is trivially true.
Let us assume that it is true for $n-1$ and make the following calculation:
\beqa
& &(\vac|\A{n}^*\ldots\A1^*\Ahh1\ldots\Ahh{n}\vac)\nonumber\\
&=&\sum_{k=1}^{n}(\vac|\A{n}^*\ldots\A2^*\Ahh1\ldots[\A1^*,\Ahh{k}]\ldots\Ahh{n}\vac)
\nonumber\\
&=&\sum_{k=1}^{n}\big\{\sum_{l=k+1}^{n}
(\vac|\A{n}^*\ldots\A2^*\Ahh1\ldots\nonumber\\
&\ldots& [[\A1^*,\Ahh{k}],\Ahh{l}]\ldots\Ahh{n}\vac)
\nonumber\\
&+&(\vac|\A{n}^*\ldots\A2^*\Ahh1\ldots\check{k}\ldots\Ahh{n}\A1^*\Ahh{k}\vac)\big\}.
\eeqa
Terms containing double commutators vanish in the limit by Lemma \ref{com} b and Lemma \ref{norm} a.
The remaining terms factorize by the preceding Proposition and by Lemma \ref{norm} b and c:
\beqa
& &\lim_{T\to\infty}(\vac|\A{n}^*\ldots\A2^*\Ahh1\ldots\check{k}\ldots\Ahh{n}\A1^*\Ahh{k}\vac) 
\nonumber\\
&=&\lim_{T\to\infty}(\vac|\A{n}^*\ldots\A2^*\Ahh1\ldots\check{k}\ldots\Ahh{n}|\vac)\cdot \nonumber\\& & \cdot(\vac|\An1^*E_m\Ahhn{k}\vac).
\qquad
\eeqa
This quantity factorizes into two-point functions by the induction 
hypothesis.\qed\\
It is also evident from the proof that the scalar product of two asymptotic
states involving different numbers of operators is zero. The Fock structure
of asymptotic states follows by standard density arguments making possible the 
usual definition of the $S$ matrix.  
\section{Conclusion}
We have constructed a scattering theory of massive particles without
the lower and upper mass gap assumptions. The Lorentz covariance of
the construction can be verified by application of standard arguments 
\cite{Ara}. Including Fermions would cause no additional difficulty, 
as the fermionic creation operators are bounded uniformly in time \cite{Buch2}.

The only remaining restriction is the regularity assumption A'. We note that it
was used only to establish the existence of scattering states - the construction 
of the Fock structure was independent of this property. Moreover, we would like
to point out that it does not seem possible to derive it from general postulates.
In fact, let us consider the generalized free field $\Phi$ with the commutator fixed
by the measure $\sigma$:
\beq
[\Phi(x),\Phi(y)]=\int d\sigma(\lambda)\Delta_{\lambda}(x-y),
\eeq 
where $\Delta_{\lambda}$ is the commutator function of the free field of mass $\sqrt{\lambda}$.  
Suppose that the measure $\sigma$ contains a discreate mass $m$ and in its neighbourhood
is defined by the function $F(\lambda)=1/\ln|\lambda-m^2|$. Then it can be
checked that every polynomial in the fields smeared with Schwartz class functions 
violates the assumption A'. However, the existence of scattering states can easily be verified
using the properties of generalized free fields. These observations indicate that 
the condition A' is only of a technical nature. To relax it one should probably look for a construction 
of asymptotic states which avoids Cook's method - perhaps similarly to the scattering theory
of massless particles \cite{Buch2,Buch4}. 

\bigskip

\noindent{\bf Acknowledgements:}
I would like to thank Prof. J. Derezinski for bringing the subject of embedded masses in the
Haag-Ruelle theory to my attention. Moreover, I am very grateful to 
Prof. D. Buchholz for inspiring discussions, numerous valuable suggestions and,
especially, for communicating the proof of Lemma \ref{sT}. I also gratefully
acknowledge stimulating discussions with Prof. S. Gierowski.

This work was partly supported by the Postdoctoral Training Program HPRN-CT-2002-0277. I also acknowledge support from the EC Research Training Network 'Quantum Spaces - Non-commutative Geometry'.

\end{document}